\documentclass[twocolumn]{aastex62}

\usepackage{graphicx}	
\usepackage{amsmath}	

\newcommand{\enzo}{\texttt{ENZO}}
\newcommand{\moray}{\texttt{MORAY}}

\newcommand{\Ms}{{\ensuremath{M_{\odot} }}}
\newcommand{\Zs}{\ensuremath{Z_\odot}}
\newcommand{\Ls}{\ensuremath{L_\odot}}

\newcommand{\K}{\ensuremath{\mathrm{K}}}
\newcommand{\gcc}{\ensuremath{\mathrm{g}\,\mathrm{cm}^{-3}}}

\newcommand{\cc}{\ensuremath{\mathrm{cm}^{-3}}}
\newcommand{\rev}[1]{\textcolor{black}{#1}}

\newcommand{\note}[1]{#1}

\graphicspath{{./}{figures/}}

\shorttitle{Pop~III Supernova Remnants}
\shortauthors{Chen et al.}

\usepackage{subfigure}

\begin{document}

\title{How Population III Supernovae \rev{Determined the Properties of the First Galaxies}}

\author{Ke-Jung Chen}
\affiliation{Institute of Astronomy and Astrophysics, Academia Sinica, Taipei 10617, Taiwan}
\email{kjchen@asiaa.sinica.edu.tw}

\author{Ching-Yao Tang}
\affiliation{Institute of Astronomy and Astrophysics, Academia Sinica, Taipei 10617, Taiwan}
\affiliation{Department of Physics, National Taiwan University, Taipei 10617, Taiwan}

\author{Daniel J. Whalen}
\affiliation{Institute of Cosmology and Gravitation, Portsmouth University, Portsmouth, UK } 

\author{Meng-Yuan Ho}
\affiliation{Institute of Astronomy and Astrophysics, Academia Sinica, Taipei 10617, Taiwan}
\affiliation{Department of Physics, National Taiwan University, Taipei 10617, Taiwan}

\author{Sung-Han Tsai}
\affiliation{Institute of Astronomy and Astrophysics, Academia Sinica, Taipei 10617, Taiwan}
\affiliation{Department of Physics, National Taiwan University, Taipei 10617, Taiwan}

\author{Po-Sheng Ou}
\affiliation{Institute of Astronomy and Astrophysics, Academia Sinica, Taipei 10617, Taiwan}
\affiliation{Department of Physics, National Taiwan University, Taipei 10617, Taiwan}

\author{Masaomi Ono}
\affiliation{Institute of Astronomy and Astrophysics, Academia Sinica, Taipei 10617, Taiwan}
\affiliation{Astrophysical Big Bang Laboratory, RIKEN, Saitama 351-0198, Japan}

\begin{abstract}

Massive Pop~III stars can die as energetic supernovae that enrich the early universe with metals and determine the properties of the first galaxies.  With masses of about $10^9$ \Ms\ \rev{at $z \gtrsim 10$}, these galaxies are believed to be the ancestors of the Milky Way. This paper investigates the impact of Pop~III supernova remnants (SNRs) from both Salpeter-like and top-heavy initial mass functions (IMFs) on the formation of \rev{first} galaxies with high-resolution radiation-hydrodynamical simulations with the \enzo\ code. Our findings indicate that SNRs from a top-heavy Pop III IMF produce more metals, leading to more efficient gas cooling and earlier Pop II star formation in the first galaxies. \rev{From a few hundred to a few thousand Pop~II stars can form in the central regions of these galaxies.} These stars have metallicities of $10^{-3}$ to $10^{-2}$, \Zs, \rev{greater than those of extremely metal-poor (EMP) stars.  Their} mass function follows a power-law distribution with $dN(M_*)/dM_* \propto M_*^{\alpha}$, where $M_*$ is stellar mass \rev{and $\alpha = 2.66 - 5.83$ and is steeper for a top-heavy IMF.  We thus find that EMP stars were not typical of most primitive galaxies.} 

\end{abstract}

\keywords{(cosmology:) early universe --- galaxies: dwarf --- galaxies: 
formation --- hydrodynamics --- methods: numerical --- stars: Population III}


\section{Introduction}
\label{Introduction}

The birth of primordial (Pop~III) stars at $z \sim 20-25$ marked the end of the cosmic dark ages and the onset of the first galaxy and supermassive black hole (SMBH) formation \citep{wa08a,wa08b,wise12,wf12,smidt18,latif21a,latif22a,pat23a}. Pop~III stars are thought to form in primordial mini halos that have grown to $10^5-10^6$ \Ms\ by mergers and accretion, when enough H$_2$ can form to cool gas and create stars. The original numerical simulations of Pop~III star formation (SF) suggested that they have typical masses of $100 - 200$ \Ms\ and form in isolation, one per halo \citep{bcl01,nu01,abn02}, but later studies have since shown that they can form in binaries \citep{turk09,sb13} or small multiples \citep{stacy10,clark11,get11,riaz18,susa19,wollenberg20,sharda20}. The final mass of Pop~III stars depends on how many of them form in a halo, when their ionizing UV flux halts accretion \citep{tm08,hos11,susa13,hir13,hir15}, and whether or not they are ejected from the disk by gravitational torques \citep{get12}. They have been found to range from less than a solar mass to up to several hundred solar masses.  However, the Pop~III IMF remains unknown because Pop~III stars cannot be directly observed, and recent simulations included crucial physics such as magnetic fields and radiation transport cannot evolve to the main sequence at the required numerical resolution \citep{fsg09,glov12,dw12,greif14,mckee20,stacy22,jaura22}.

Pop~III stars radically transform their environments, first by photoevaporating the halos in which they are born \citep{wan04,ket04,awb07} and at least partially ionizing others nearby \citep{su06,hus09}.  Then, depending on their masses, some explode as supernovae (SNe), expelling large masses of the first metals in the universe.  Pop~III stars with masses of $8 - 30 \Ms$ die as core-collapse (CC) SNe and $90 - 260$ \Ms\ stars explode as highly energetic pair-instability (PI) SNe \citep{hw02,hw10}.  A few very rapidly-rotating $30-60$ \Ms\ Pop~III stars may die as gamma-ray bursts \citep[GRBs; e.g.,][]{mes13a} or hypernovae \citep[HNe; e.g.,][]{smidt13a, chen17c}.   A number of studies have examined how metals from Pop~III SNe propagate into the universe on a variety of scales, from inside the halo itself \citep{slud16} to out into the relic H II region of the progenitor star \citep{get07,ritt12,rit16,chen17b,mag20,taru20,latif20c}.  Both radiation \citep{yoh07,suh09} and metals \citep{mbh03,schn06,ss07,brit15} from the first stars cause subsequent generations of stars to form on smaller mass scales.  Pop~III SNe could constrain the masses of the first stars, either directly through their detection in the near-infrared \citep[NIR;][]{tet12,tet13,ds13,wet12a,wet12c,wet13c,wet12b,ds14,hart18a,moriya19,ryd20a} or indirectly from their nucleosynthetic imprint on less massive second-generation stars that may still live today \citep[e.g.,][]{fet05,bc05,iet05,jet09b,frebel10,jw11,susa14,chen17a,chen17b,ish18,hart18b,hart19a,chiaki18,chiaki20}.

When the halos grow to $10^8 - 10^9$ \Ms\ through hierarchical structure formation by mergers and accretion, they are massive enough to maintain consecutive cycles of stellar birth and explosion without all the fuel for forming new stars being blown into the IGM, becoming, in effect, the first galaxies \citep{get08,fg11}. \rev{Unlike other low-mass dwarf galaxies across cosmic time, the baryon components of these first galaxies were primordial gas and Pop~III SNRs.} Ionizing photons, winds, and SNe from stars in these galaxies regulate the rise of later generations and determine their observational signatures \citep{get10,jeon11,pmb12,wise12,ss13,jeon14,corl18,jeon19a,abe21}.  The simulations of first galaxies performed to date follow their formation in cosmological environments from early times but assume a fixed IMF for Pop~III and Pop~II stars that does not evolve over time \citep[e.g.,][]{ren15,fiby15}. 

To probe the Pop~III IMF, we take Pop~III SNRs from two IMFs, Salpeter-like \citep{sal55} and Hirano \citep{hir15}  as an initial condition and simulate the evolution of SNRs with the primordial gas until the first galaxies form. Our goal is to understand how the properties of the first galaxies vary with the properties of Pop~III SNRs that depend on the Pop~III IMF. The observational signatures of these galaxies may be revealed by the {\em James Webb Space Telescope} \citep[{\em JWST};][]{jwst,jwst2} and $30 - 40$ m telescopes on the ground in the coming decade. 

The structure of this paper is as follows. In Section 2, we describe our numerical methods and protogalaxy models.  We present the simulations of the collapsing Pop~III SNRs in the proto-galaxy in Section 3, and the resulting Pop~II SF in Section 4. Finally, we discuss the limits of our simulations in Section 6 and conclude in Section 7.

\section{Numerical Method}
\label{Methods}

We model the formation of first galaxies with the \enzo\ adaptive mesh refinement (AMR) cosmology code \citep[v2.5;][]{enzo}.  \enzo\ uses an adaptive particle-mesh $N-$body scheme \citep{efs85,couch91} to evolve dark matter (DM) and a third-order accurate piecewise-parabolic method for gas flow \citep{wc84,bryan95}.  We use the low-viscosity Harten-Lax-van Leer-Contact (HLLC) Riemann method \citep{toro94} for capturing strong shocks and rarefaction waves to prevent negative energies or densities in the simulation.  \enzo\ self-consistently evolves nine-species non-equilibrium gas chemistry with hydrodynamics \citep[H, H$^+$, e$^-$, He, He$^+$, He$^{++}$, H$^-$, H$_2$ and H$_2^+$;][]{abet97,anet97} and includes primordial gas cooling in the energy equation:  collisional excitational and ionizational cooling by H and He, recombinational cooling, bremsstrahlung cooling and H$_2$ cooling.  We use H$_2$ cooling rates from \citet{ga08} and metal cooling rates from \citet{japp07}.  

\subsection{Protogalaxy Model}

The protogalaxies in our simulations are approximated as isolated Navarro-Frenk-White (NFW) halos \citep{nfw96} \rev{whose DM density profiles are
\begin{equation}
  \rho_{\rm DM}(r) = \frac{\rho_0}{\frac{r}{R_{\rm s}}(1+\frac{r}{R_{\rm s}})^2},   
\end{equation}
where $r$ is the radius and $\rho_0$ and the "scale radius" $R_\mathrm{c}$ depend on the physical properties of the halo.} This profile is used to initialize the gravitational potential of the DM in the halo but DM dynamics itself is not included in our runs so this potential held fixed over the run. We adopt idealized profiles for the initial structure of the protogalaxy instead of evolving them from primordial density fluctuations at high redshifts so we can efficiently run models and examine the impact of Pop~III SN remnants (SNRs) on their structures at later times.  Because the DM halos for first galaxies formed at $z \sim 10 - 15$, these halos have a more compact structure than our present-day galaxies. For our simulations, we use the scale radius of 1000 pc for a non-rotating $10^9$ \Ms\ DM halo. with a total gas mass of $M_\mathrm{g} \sim 1 \times 10^8$ \Ms\ for our protogalaxy.

\rev{DM in the first galaxies is composed of swarms of DM minihalos that have gravitationally congregated into a more massive bound structure, not the smooth NFW profiles with which we initialized our models.  Most of the mass in the galaxy is in the form of these smaller halos. Our simulations therefore do not properly account for their point-mass contribution to the true potential in the halo.  The immediate consequence of this for our models is that they underestimate mixing of metals from SNRs in the interior of the halo.  Tidal torquing between constituent minihalos in the galaxy would eject streams of gas from their SNRs and promote mixing of their metals with ambient pristine gas.}

\begin{figure}
	\centering
	\includegraphics[width=\columnwidth]{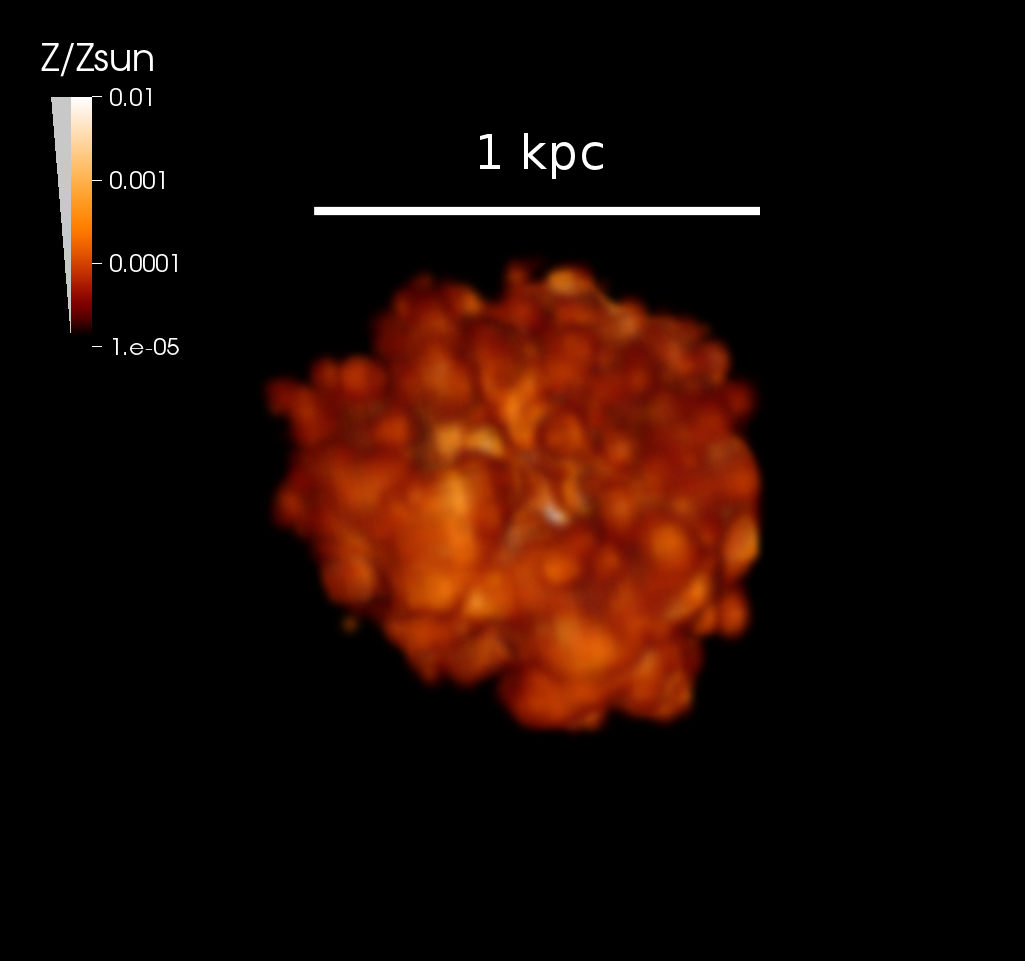}
	\caption{Metallicity distribution of a Pop~III SNR.  \rev{This iso-surface plot shows the morphology of a HN SNR of a 60 \Ms\ Pop~III star 10 Myr after the explosion}. \rev{The SNR has expanded to a radius of 1 kpc and enriched the surrounding gas to metallicities of $\sim 10^{-4} - 10^{-2}$ \Zs\ \citep{chen15}.}}
	\label{fig:snr}
\end{figure}

\begin{figure*}
\centering
\includegraphics[width=\textwidth]{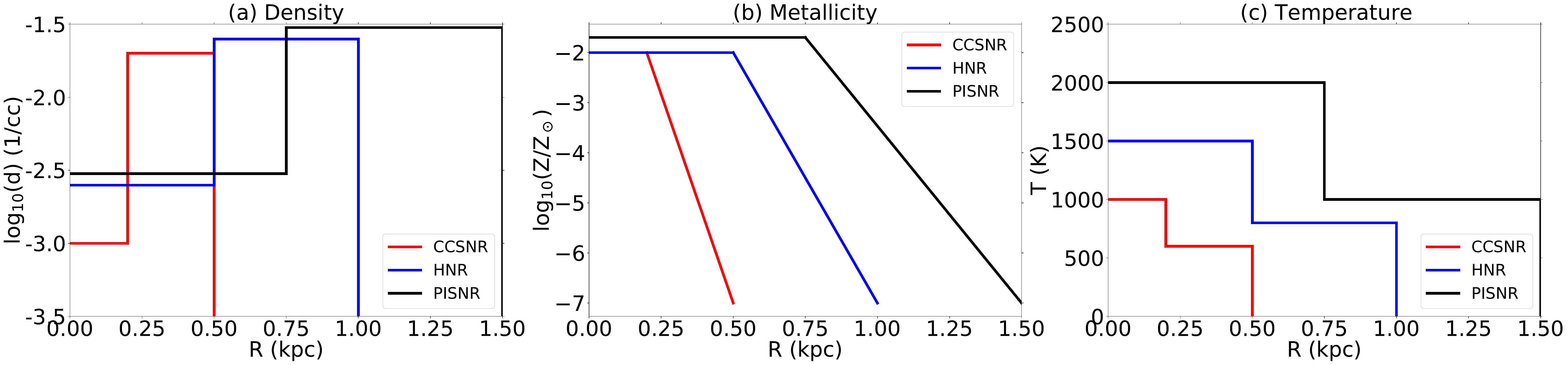}
\caption{Left to right: radial profiles of number density, metallicity, and temperature of the three types of SNRs in our simulations, respectively.}
\label{fig:snrprofile}
\end{figure*}

Pop~III SNRs are initialized from Pop~III SNe in cosmological environments in \citet{chen14c, chen15}, as shown in Figure~\ref{fig:snr}. Depending on the SN type, Pop~III SNRs can reach radii of $0.5 - 2$ kpc with metallicities of $\sim 10^{-4} - 10^{-2}$ \Zs. \note{By assuming spherical symmetry and neglecting the clumpy structures, we calculate 1D simplified profiles of density, temperature, and metallicity of Pop~III SNRs based on \citet{chen15} and show these profiles in Figure~\ref{fig:snrprofile} and initialize these profiles as the pre-existing SNRs at the beginning of each run.}  Based on \citet{chen15}, these profiles consider the relic H II region of the star, which can expand the size of SNRs after it comes into pressure equilibrium in the warm, partially recombined gas \citep{get07}.  Our SNRs are produced by three types of SN explosions: a 15 \Ms\ CC SN (1.2 foe), a 60 \Ms\ HN (10 foe), and a 250 \Ms\ PI SN (100 foe), where 1 foe $=$ 10$^{51}$ erg.  The metal yields of these explosions are 1.8 \Ms, 26.4 \Ms, and 109 \Ms, respectively.  The SNRs are randomly distributed throughout the halo but are all assigned free-fall velocities towards its center.  

Because we do not evolve halos in cosmological environments, we ignore temperature floors in the gas imposed by the cosmic microwave background (CMB).  \rev{The CMB can affect the mass scales of fragmentation in gas in minihalos at early times.  However, at the typical redshifts of formation of first galaxies the CMB floor is 30 - 45 K, similar to the temperatures to which metals and dust can cool gas at the low metallicities in our models.  Exclusion of the CMB will therefore have negligible effects on the mass scales of star formation in our simulations.}

Our $10^9$ \Ms\ DM halo is composed of hundreds of minihalos but only a few tens of them were originally massive enough to host Pop III stars or SNRs.  The initial numbers and types of SNRs in the massive halo are determined by the choice of Pop~III IMF.  Here, we consider a Salpeter-like IMF \citep[SAL;][]{sal55, kroimf, chaimf} with a peak at 10 \Ms\ and a top-heavy IMF from \citet{hir15} (HIR15).  \rev{We assume a Salpeter IMF with a peak at 10 \Ms\ for  massive stars because it is the approximate lower limit in progenitor mass for CC SNe.  A number of recent studies also suggest that Pop III stars may have had typical masses of a few tens of solar masses \citep{hus09,hos11,hir13,hir15,latif22a}.}  

If we use the SAL IMF, 18 CC~SNe and 2 HNe are initialized in the halo. In the case of the HIR15 IMF, the halo contains 2 CC~SNe and 7 PI~SNe.  We randomly distribute these SNRs throughout the halo.  Gas densities in our halos marginally trace the NFW DM profiles because the gas is assumed to have already settled in the gravitational potential to some degree by the time we begin to evolve them. We consider two gas density profiles of $\rho(r) \sim r^{-1}$ and $\rho(r) \sim r^{0}$ in the halo to examine the effect of gas distribution formed during the assembly of mini halos. Furthermore, all density profiles are non-hydrostatic equilibrium.

The halo is centered in a 10 kpc box with a 128$^3$ root grid with outflow boundary conditions. To resolve the structure of SNRs, we use up to 10 levels of AMR. The grid is flagged for refinement where the gas overdensity is 8$\rho_0 N^l$, where $\rho_0 =$ 1.673 $\times$ 10$^{-27}$ \gcc\ is the ambient density of the halo, $N =$ 8 is the refinement factor, and $l$ is the AMR level of the grid patch. In the densest regions, this procedure yields a maximum spatial resolution of 0.077 pc, which is sufficient to resolve the shell mergers between SNRs, and the central SF sites at later times.

\subsection{Star Formation}
\label{SF}

Our galaxy simulations have a spatial resolution of up to 0.077 pc, which is high enough to resolve metal mixing and star-forming clouds in the halo at later times but not Pop~II SF.  
\rev{In the high density region, each cell mass represents a small patch of the molecular cloud that later forms a single star.} We adopt a subgrid model of SF from \cite{co92} and assume a Pop~II SF efficiency $\epsilon_\mathrm{SF} =$ 0.05, in which 5\% of the mass of the gas in a cell that satisfies the criteria for SF is converted into a star particle. This conversion efficiency is motivated by studies of SF in the local universe that suggest that only a small fraction of molecular gas is converted into stars \citep{evans09,kru19,kim21}. In reality, it may vary depending on the resolution of the simulations and the physical conditions in SF regions. Since cloud structure is marginally resolved in our simulation, the SF efficiency here represents a lower limit. \rev{SF cells in our simulations normally contain $\sim 20 - 500$ \Ms\ of gas, which leads to typical masses of several to tens of solar masses for Pop II stars.}    

\subsection{Stellar Feedback}
\label{StellarFeedback}

After star particles form in the simulations, their radiation feedback to surroundings is on during their lifetime. We model the feedback of ionizing UV photons from massive stars with the \moray\ ray-tracing package \citep{moray}. \moray\ includes radiation pressure on gas due to photoionization and the radiation is self-consistently coupled to gas hydrodynamics, cooling, and chemistry in \enzo. Each star particle is treated as a point source that emits both ionizing and Lyman-Werner (LW) photons. The luminosities and lifetimes for new-born Pop~III stars\footnote{\note{These newborn Pop~III stars known as Pop~III.2 \citep{hir15} might have experienced the radiative feedback from earlier Pop~III stars. Since we do not include the radiative feedback from these Pop~III stars, we treat the Pop~III.2 stars as Pop~III stars.}} formed during the simulations use the table from \citet{s02}. For individual Pop~II stars, we have calculated their luminosity function using 1D stellar evolution models with \texttt{MESA} \citep{paxt11,paxt13,paxt15,paxt18}.  

Stellar winds may reshape the ambient media of massive stars and affect the dynamical evolution of SNRs \citep{dw05}. However,
we neglect the feedback of stellar winds from Pop~III stars because the zero metallicity suppresses the stellar wind significantly. On the other hand, stellar wind of Pop~II stars vary as $\dot{M} \propto Z^{m}$ where $m \sim 0.1 - 1$ for metallicities down to 0.01 \Zs\ \citep{vink01,nsmith14}. Since the Pop~II stars in the first galaxies likely form with metallicities of  $Z < 0.01 \Zs$, the winds for these low-metallicity stars are weak and uncertain \citep{ou22}. Therefore, we also exclude the wind feedback of Pop~II stars in our runs.  

Some of the massive Pop~II and Pop~III stars eventually die as SNe at the end of their life and produce strong feedback to their surroundings. Explosion energy from Pop~II and Pop~III SNe is deposited in the gas as thermal energy rather than linear momentum \citep[e.g.,][]{wet08a}.  This practice in principle can lead to the classic overcooling problem, in which large amounts of thermal energy deposited in high densities are radiated away by cooling before they can create the large pressure gradients that drive shocks outward, as in real explosions.  \rev{However, this issue is resolved in our runs because our high resolution allows us to follow radiative cooling accurately.} Furthermore, as discussed in the Introduction \ref{Introduction}, ionizing UV flux from the star drives gas from it in strong supersonic flows and the explosions always occur in low densities \citep[stellar winds have a similar effect when present;][]{chen15, smidt18}.  Consequently, sound-crossing times in our models are shorter than cooling times and SNe can drive strong shocks in our runs.

We performed eight runs in which we used the initial SNRs based on two Pop~III IMFs, and two mass scales of the Pop~III stars that form during the run, and two gas density profiles for the halo.  These models are divided into four groups, and what varies between models in a group is the mass scale of Pop~III stars that form in them.  The model parameters are summarized in Table~\ref{table:simic}.  SF rates in the galaxies in our models level off as ionizing UV, and SNe due to stars begin to suppress the formation of new stars.  Since we are mostly interested in the transition from Pop~III to Pop~II SF (and computational costs rise as more stars form), the simulations are evolved for $35 - 60$ Myr when SF flattens out and the properties of the stellar populations in the nascent galaxies stabilize.

\begin{table}
	\centering
	\begin{tabular}{c c c c c c c c }
		Model & $M_\mathrm{halo}$ & SNR IMF & $\rho(r)$ & 
		$M_\mathrm{PopIII}$ & $N_\mathrm{CC}$ & $N_\mathrm{HN}$ & $N_\mathrm{PI}$\\
		\hline
		A1    &   $10^9$   &   SAL       &   $r^{-1.0}$   &   20    &   18  &   2   &   0   \\
		A2    &   $10^9$   &   SAL       &   $r^{-1.0}$   &   200  &   18  &   2   &   0   \\
		\hline
		B1    &   $10^9$   &   HIR15   &   $r^{-1.0}$   &   20    &   2    &   0   &   7   \\
		B2    &   $10^9$   &   HIR15   &   $r^{-1.0}$   &   200  &   2    &   0   &   7   \\
		\hline
		C1    &   $10^9$   &   SAL       &   $r^0$         &   20    &   18  &   2   &   0   \\
		C2    &   $10^9$   &   SAL       &   $r^0$         &   200  &   18  &   2   &   0   \\
		\hline
		D1    &   $10^9$   &   HIR15   &   $r^0$          &   20    &   2    &   0   &   7   \\
		D2    &   $10^9$  &   HIR15   &   $r^0$          &   200  &   2    &   0   &   7   \\
		\hline
	\end{tabular}
	\caption{Model parameters. Left to right: model, halo mass (\Ms), Pop~III SNR IMF, baryon density profile, Pop~III star mass (\Ms), number of CC~SNRs, number of HN~SNRs, and number of PI~SNRs.}
	\label{table:simic}
\end{table}

\section{Accretion of Population III Supernovae Remnants}
\label{Results}

\subsection{Collapse of Pop~III SNRs}

\begin{figure*} 
	\centering
	\includegraphics[width=\textwidth]{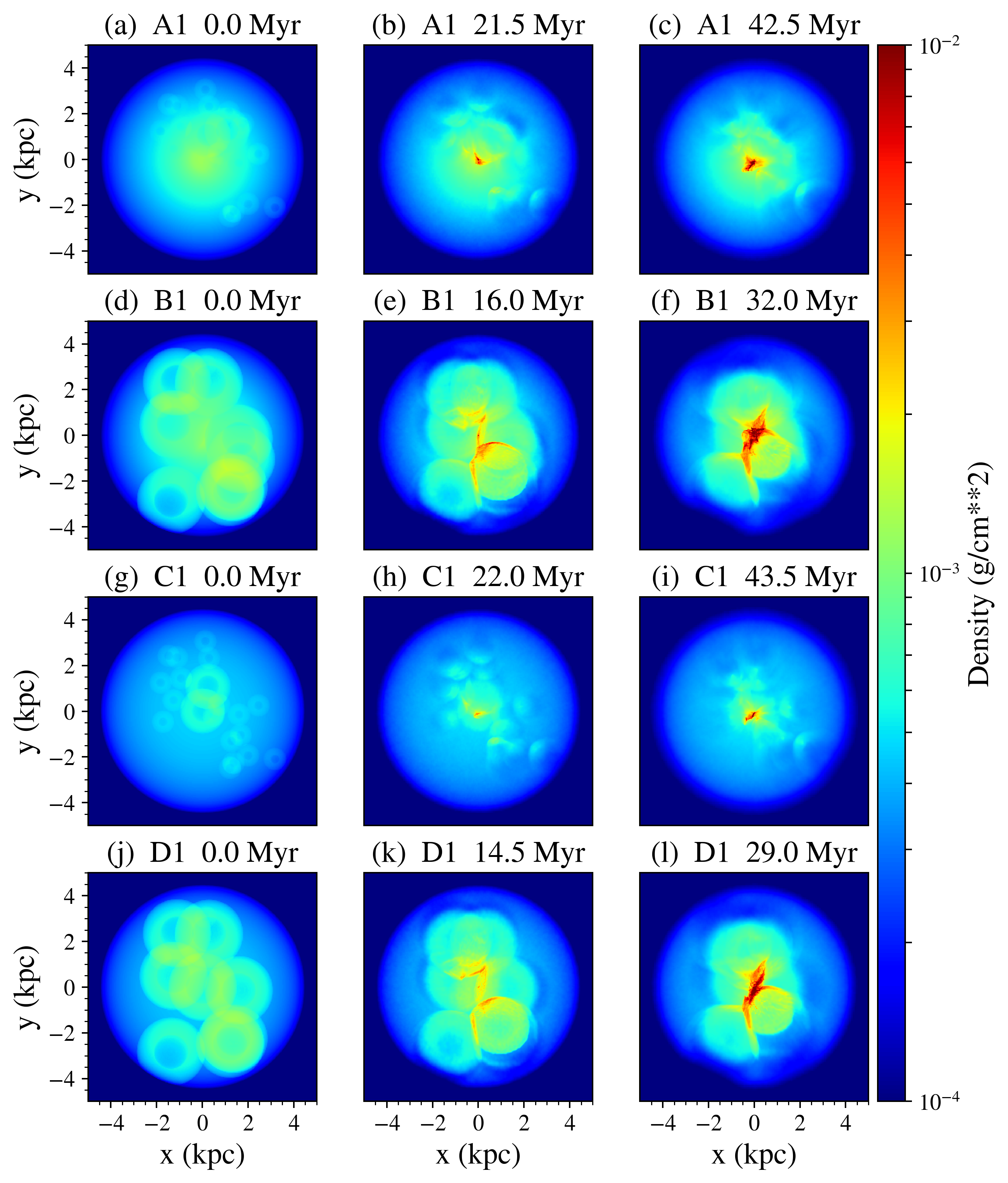}
	\caption{ Density distribution at the beginning, intermediate, and final stages of the A1, B1, C1, and D1 runs before the Pop~II SF. Collisions of Pop~III SNRs create filamentary structures and drive the mixing between the metals and primordial gas. The mixed gas eventually collapses to the halo center, where SF would occur shortly.}
	\label{fig:accgas}
\end{figure*}

The in-falling time for the outermost SNRs to reach the halo center is about $30 - 40$ Myr. Since the gravity of the halo is dominated by the dark matter, the in-falling time for most of gas to reach the center is similar among models. 
We first show the density evolution of simulations in Figure~\ref{fig:accgas}. At $t=0$, the range of filling factors of SNRs is due to the energy of SN explosions: larger PI SNRs in B1 and D1 are easily distinguished from the less energetic CC~SNRs in the A1 and C1 runs. These Pop~III SNRs and the primordial gas are dragged by the halo gravity toward its center. Meanwhile, the SNR-SNR collisions produce a range of turbulent flows at the center, from fairly violent ones in the B and D models. Eventually, the turbulence drives the mixing of SNR metal and primordial gas and creates filamentary structures that soon form into dense clumps due to the self-gravity and metal cooling of the gas.
 
\subsection{Feedback from New Population III Supernovae }

Before the initial SNRs accrete, the primordial gas has been accumulated at the halo center and formed Pop~III stars, which compositions remain primordial but experience the UV radiation from earlier Pop~III stars. These Pop~III stars only form in the A and B runs with a steeper gas density profile of $r^{-1}$. For C and D runs, the pre-exiting SNRs around the halo center prevent the consequent Pop~III SF. 

\begin{figure*}
\centering
\includegraphics[width=0.8\textwidth]{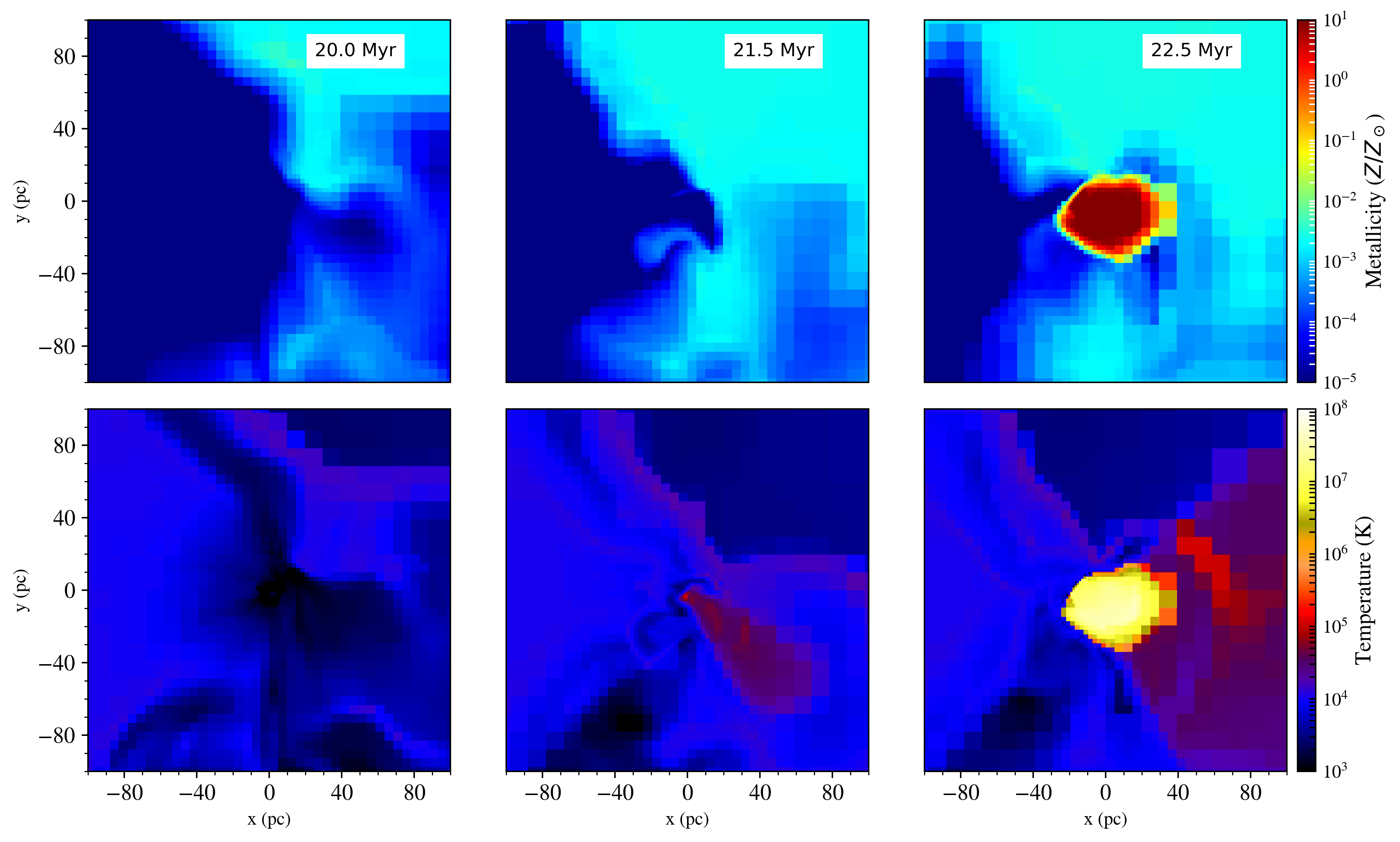}
\caption{Metallicity and temperature slices of the central 200 pc, where a 200 \Ms\ Pop~III star forms in A2.  Left to right: 20 Myr (right before birth), 21.5 Myr (1.5 Myr after birth), and 22.5 (0.5 Myr after death).  After the star born, its radiation heats and ionizes the surrounding gas and creates a giant HII region as shown in the middle panel.  When the star dies as a PI SN, the explosion forms a new SNR of hot and metal-rich ejecta visible in the right panels.}
\label{fig:firstsne}
\end{figure*}

These massive Pop~III stars can impose strong radiative and SN feedback before the initial Pop~III SNRs reach the halo center. 
We show radiative feedback and metal injection of a Pop~III star in Figure~\ref{fig:firstsne}.  This Pop~III star of 200 \Ms\ heats and ionizes surrounding gas, which can either suppress or promote new SF in its vicinity \citep{wet08b,wet10}. After its short lifetime of $\sim 2.0$ Myr, the star dies as a PI SN and its shock heats the gas to high temperatures ($> 10^{5}$ K) and ejects a large mass of metals that enhance cooling and promotes a transition to Pop~II SF.  The existence of Pop~III stars in the first galaxies is promising, but due to a rapid pollution of metals from Pop III SNRs; only a small fraction of pristine gas could form Pop~III stars. Therefore, Pop~III stars cannot be the major components of the first galaxies. Only one or two Pop~III stars could form in A and B models; their metal contributions to the halo is small in compared to that of the initial Pop~III SNRs. However, most of Pop~III stars form around the halo center that also hosts the later Pop~II SF. Therefore, the stellar and SN feedback of Pop~III stars may still affect the Pop~II SF in the first galaxies. 

\subsection{Chemical Enrichment}

\begin{figure*}
	\centering
	\includegraphics[width=0.52\textwidth]{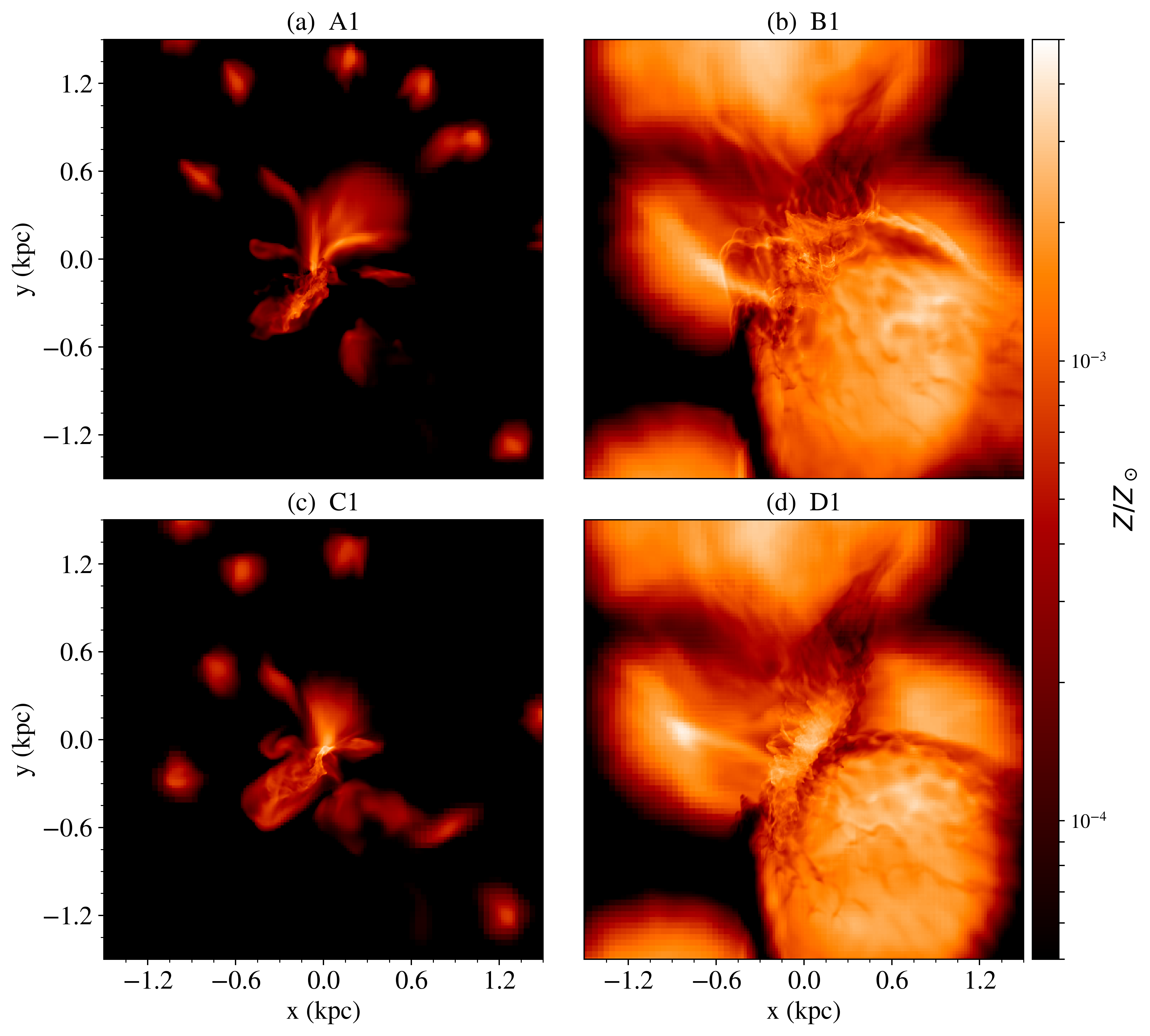} 
    \includegraphics[width=0.46\textwidth]{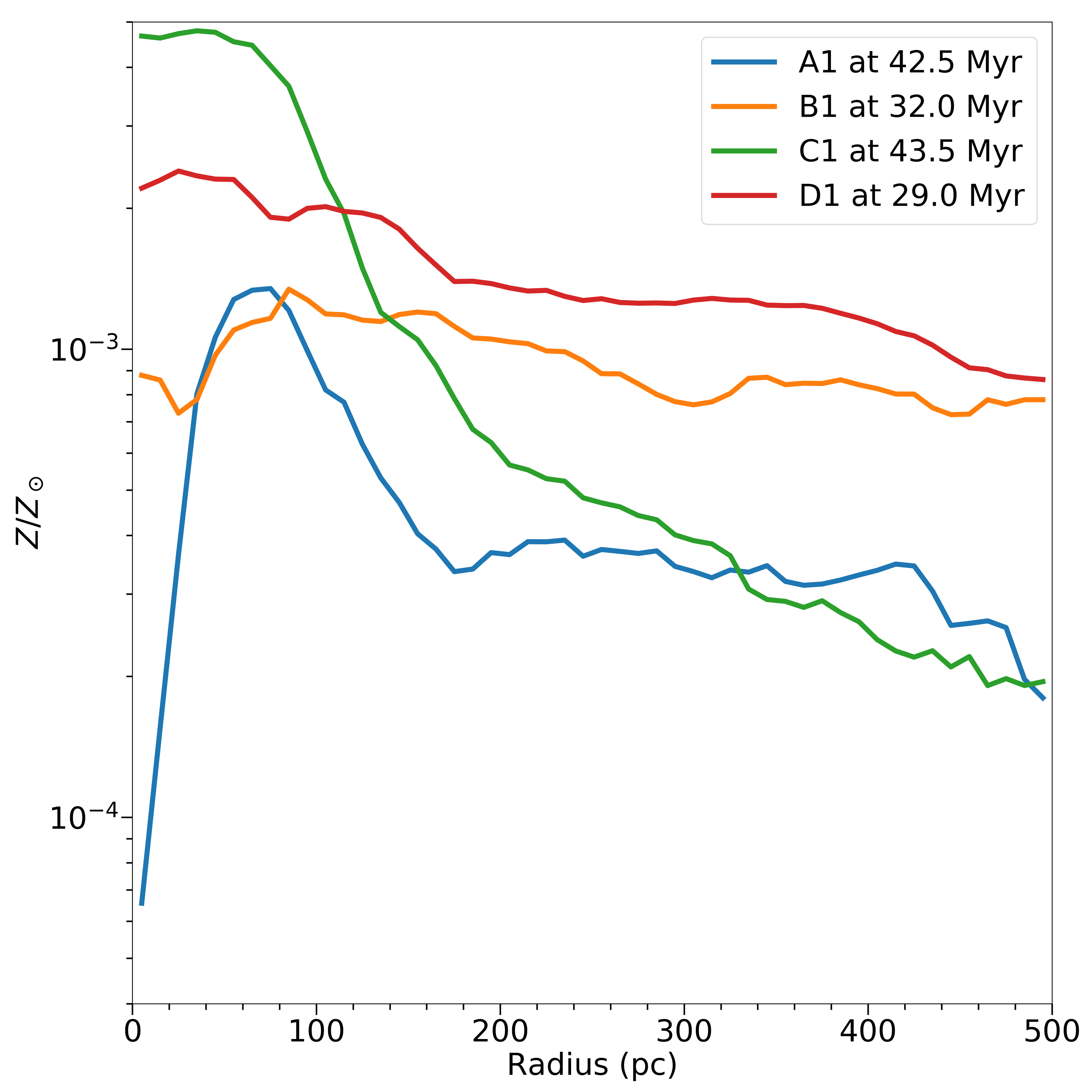}
	\caption{Left: Metallicity distribution before the Pop~II SF at A1, B1, C1, and D1 runs. Panels show slices of metallicity at the central 3.0~kpc region, where metals of the Pop~III SNRs have reached the centers and chemically enriched the primordial gas to metallicity of $10^{-4} - 10^{-3} \Zs$ at large. Right: The corresponding 1D spherically-averaged metallicity profiles of $r < 500$ pc.  }
	\label{fig:rgaszooma}
\end{figure*}

To examine the chemical enrichment by Pop~III SNRs, we first show the metallicity distribution for A1 - D1 model in the left panel of Figure~\ref{fig:rgaszooma}, before the Pop~II SF. The accretion of PI SNRs in the  HIR15 IMF cases (B1 and D1) results in much concentrated and high metallicity regions than in SAL IMF cases (A1 and C1). Despite some SNRs are still approaching to the halo center, the central metallicity of $r<200$ pc is enriched by both the new Pop~III SNRs and the original Pop~III SNRs.  We find that halos with Pop~III PI SN remnants in B1 and D1 exhibit stronger turbulent mixing in their densest regions by showing a uniform metallicity across the inner region, because the remnants collide with greater energy when they reach the halo center.  In fact, the yields of Pop~III CCSNe, HNe, and PISNe produce unique abundance patterns that can leave fingerprints on the Pop~II stars for decoding the nature of Pop~III stars. Tracing each individual element from each SN requires a huge advection network that is computationally expensive. Therefore, we do not trace the individual elements in our simulations but only consider the total amount of metal by summing all elements of SN yields.

\begin{figure*}
	\centering
	\includegraphics[width=0.52\textwidth]{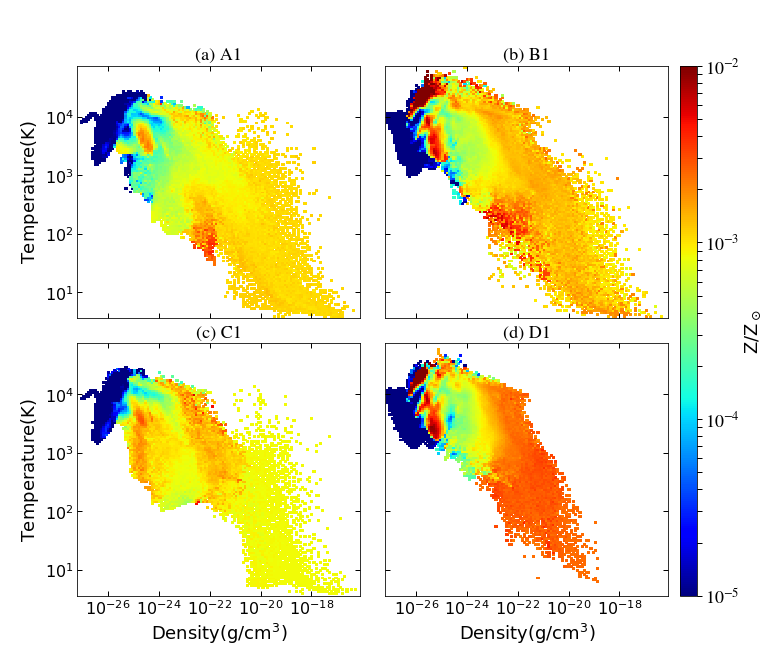} 
    \includegraphics[width=0.46\textwidth]{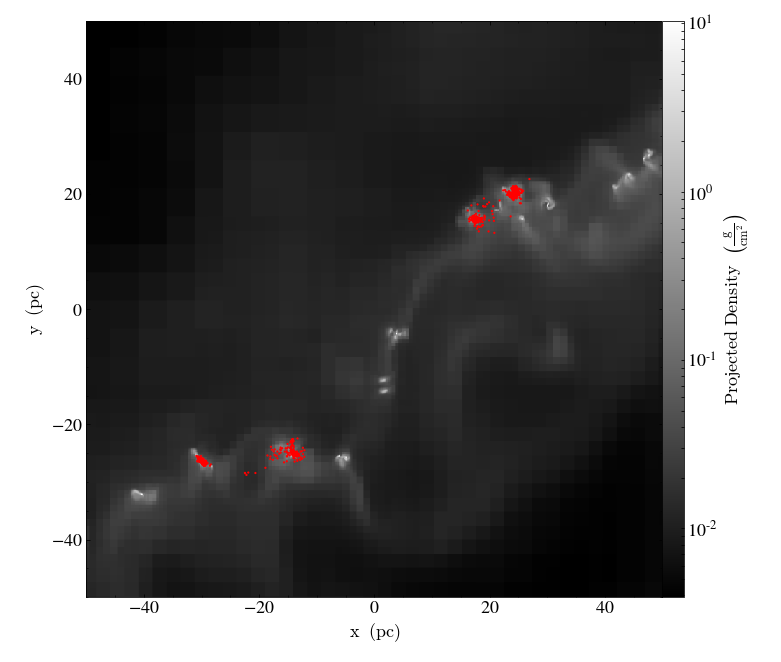}
	\caption{Left: The phase diagram of densities, temperatures, and metallicities of the gas inside the first galaxy before the Pop~II SF. The star-forming gas is located at the high-density and low-temperature region as shown in the bottom-right corn in each panel.  The gas temperatures and densities are also associated with its metallicity. The Pop~II SF region in B1 and D1 contains higher metallicity than that in A1 and C1.  \note{Some high-density regions have been cooled down to $\sim 10$ K because we ignore the ambient gas temperatures heated by the cosmic background radiation.} Right: The Pop~II SF regions in A1. Red dots represent individual Pop~II stars. Due to the metal cooling and turbulence, these Pop~II stars form into clusters along the dense filaments around the halo center.  }
	\label{fig:pop2}
\end{figure*}

We show spherically-averaged profiles of metallicities for the A1 - D1 runs before the Pop~II SF in Figure~\ref{fig:rgaszooma}.  The metallicity in all models ranges from  $2 \times 10^{-4} - 5 \times 10^{-3} \Zs$. The metallicity rises in the inner region of A1 is due to the chemical enrichment of Pop~III stars and the rise in C1 is due to an initialed SNR at the center. Finally, we show the density, temperature, and metallicity phase diagram in the left panel of Figure~ \ref{fig:pop2}. Maximum densities of the gas clumps can reach above $10^{5} \cc$ with metallicities of $Z \geq 10^{-3}$ at temperatures of $10-100$ K. As time evolves, these dense gas clumps eventually collapse and give birth to the Pop~II stars. 

\section{Pop~II Star Formation in the Protogalaxy}

The accreting Pop~III SNRs and the primordial gas eventually trigger the Pop~II SF around the halo center at $30 - 35$ Myr when the dense clumps satisfy the SF criteria.  Due to the effective metal cooling, the mass scale of these Pop~II stars shifted to a low mass-end and formed in a cluster as shown in the right panel of Figure~\ref{fig:pop2}. \note{In our simulations, the metal cooling acts on the cloud scale, and the inhomogeneous mixing of metal drives the fragmentation of clouds, forming into SF clumps. We exclude the effect of SF disk fragmentation \citep{clark11, chiaki18, sharda20} due to metal and dust cooling in the sub-grid scale.} Several star-forming regions appear in our simulations due to the inhomogeneous mixing of SNRs and primordial gas. The SF mainly locates along the filaments of dense structures within the central 200 pc of the halo. Most stars form within a few million years until their stellar feedback disrupts the star-forming could and prevents further SF. The Pop~II SF begins later in SAL halos than in HIR15 halos because a weaker metal cooling delays the SF.  Among all models, about $170 - 2,295$ Pop~II stars form in this episode of SF and produce total stellar mass  $\sim 225 - 3,966 \Ms$ (see Table \ref{table:rmainresults} for details). 

\subsection{The Mass Function}

\begin{figure*}
	\centering
	\includegraphics[width=\textwidth]{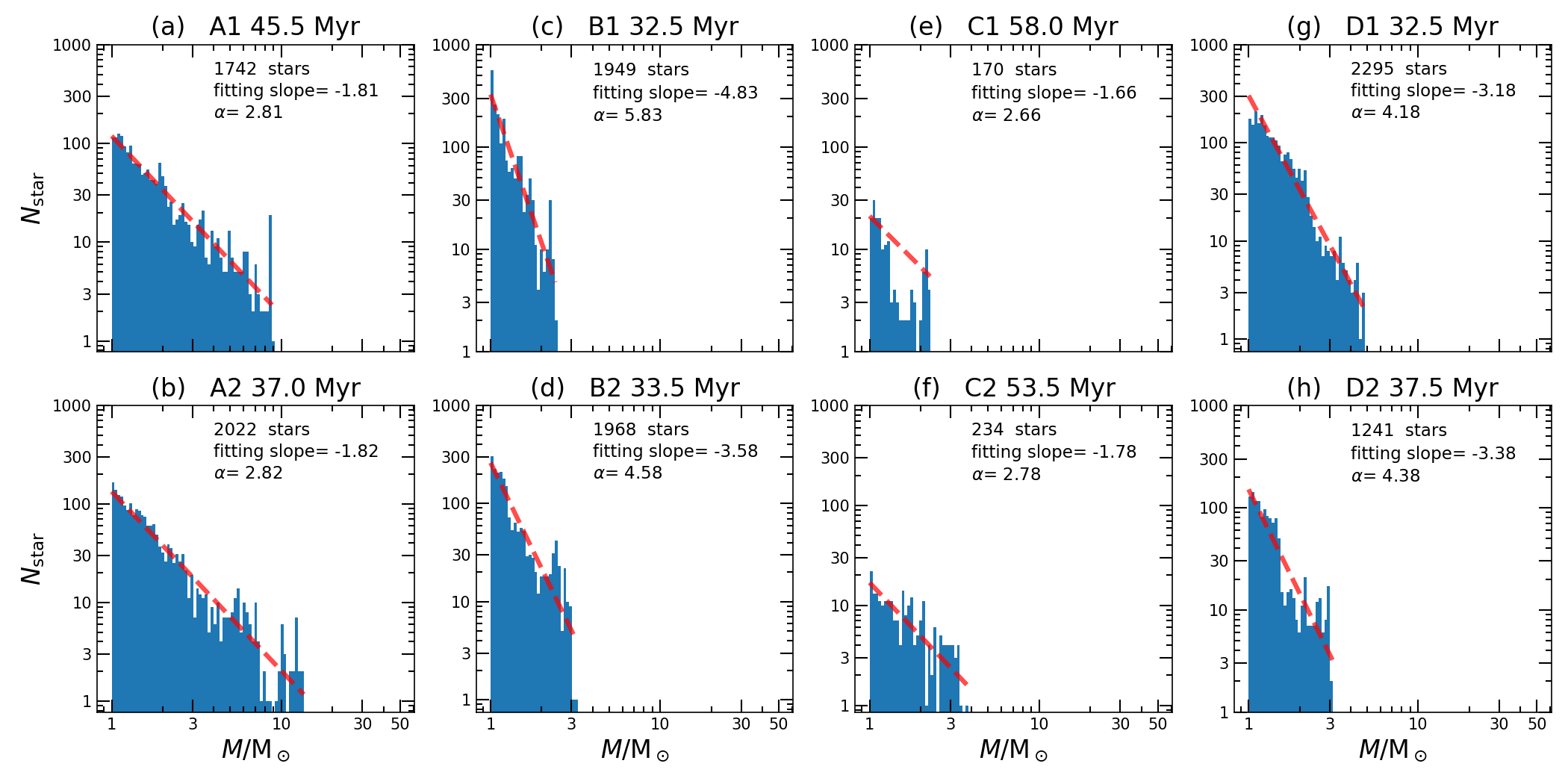}
	\caption{The Pop~II mass function in first galaxies. No Pop~III stars remain at the end of any of the runs. The mass function can be fitted with a power-law function in most runs. The total number of stars, the best-fit index $\alpha$,  and the fitting line slope 1 - $\alpha$ are listed in each panel. }
	\label{fig:rsm}
\end{figure*}

We show the mass functions of Pop~II stars in Figure~\ref{fig:rsm}. Most of the mass functions can be fitted using a power-law distribution, $dN_*/dM_* \propto M_*^{-\alpha}$, where $N_*$ is the number of stars and $M_*$ is the mass of the stars.  The most massive Pop~II star in our simulations is $\sim 13.6$ \Ms\ in A2.  In C and D models, the mass function becomes slightly irregular, with a second peak appearing at the high mass end.  In these cases, low-mass stars formed first, creating the first peak.  Then radiative feedback from these Pop~II stars heated the surrounding clouds and increased their Jeans masses, that led to more massive star formation shown as the second peaks in Figures~\ref{fig:rsm}.  These mass functions could place constraints on the Pop~III SNRs if the power-law indices of the stellar mass functions of our model galaxies hold until their Pop~II SF becomes a steady state.

We find that power-law index, $\alpha$ varies from 2.66 -- 5.83 across all eight models and this value is higher than 2.3, $\alpha$ of the Salpeter IMF \citep{sal55,kroimf,chaimf}.  Comparing all models, we find that the models seeded with PI SNRs results have larger $\alpha$ and steeper slopes in the power-law for the mass functions. Before Pop~II SF, stronger metal cooling and turbulence driven by PI SNRs can fragment enriched clouds and decrease the mass scale of Pop~II stars and result in steeper slopes. \rev{The change in IMF slope can be understood by considering the Jeans mass of the clouds, 
\begin{equation}  
M_j \approx 13.7 \Ms (\frac{T}{50\, \K})^{3/2} (\frac{\rho}{10^{-20}\,\gcc})^{-1/2},  
\end{equation}
where $\rho$ is gas density and $T$ is the temperature. The Jeans mass is more sensitive to changes in temperature than density. As shown in the phase diagram in Figure \ref{fig:pop2}, the Jeans mass varies from 1.22 - 38.7 \Ms\ for temperatures of 10 -  100 K at densities of $10^{-20} \gcc$. The cell masses are 20 to 500 \Ms, indicating that the majority of them are poised for imminent collapse and star formation. In Figure \ref{fig:pop2}, gas temperatures at $\rho \sim 10^{-20} - 10^{-18}$ \gcc\ are generally lower in B1 than in A1 because B1 has a higher metallicity. Enrichment by PI SNRs can create higher metallicities and stronger turbulence, facilitating the mixing of metals toward the center of the halo where star formation occurs. The higher metallicities cool gas to lower temperatures and result in smaller Jeans masses, promoting low-mass star formation and the steepening the IMF slope.}   

\begin{table*}
 \centering
	\begin{tabular}{c c c c c c c c c}
		Model & $t_\mathrm{evol}$ (Myr) & $N_\mathrm{PopII}$ & $M_\mathrm{*}$ ($M_\odot$) & $N_\mathrm{PopIII}$ & $\alpha$  & L (\Ls)\\ 
		\hline
		A1   &   45.5     &   1742     &   3350     &   2   &   2.81 & $3.23 \times 10^5$ \\
		A2   &   37.0     &   2022     &   3966     &   1   &   2.82 & $6.00 \times 10^5$  \\
		\hline
		B1   &   32.5     &   1949     &   2387     &   1   &   5.83 & $1.39 \times 10^4$ \\
		B2   &   33.5     &   1968     &   2655     &   1   &   4.58 & $2.76 \times 10^4$ \\
		\hline
		C1   &   58.0     &   170       &   225       &   0   &   2.66 & $1.92 \times 10^3$  \\
		C2   &   53.5     &   234       &   391       &   0   &   2.78 & $8.57 \times 10^3$ \\
		\hline
		D1   &   32.5     &   2295     &   3409     &   0   &   4.18  & $5.57 \times 10^4$ \\
		D2   &   37.5     &   1241     &   1692     &   0   &   4.38  & $1.77 \times 10^4$ \\
		\hline
	\end{tabular}
    \caption{Summary of stellar populations in all models.  Left to right:  model name, evolution time, the final number of Pop~II stars, total stellar mass, number of Pop~III stars formed during the simulation,  $\alpha$ (the power-law index of mass function, $dN_*/dM_* \propto M_*^{-\alpha}$), and total luminosity.} 
    \label{table:rmainresults}
\end{table*}

\subsection{Stellar Metallicity}

\begin{figure}
	\centering
	\includegraphics[width=\columnwidth]{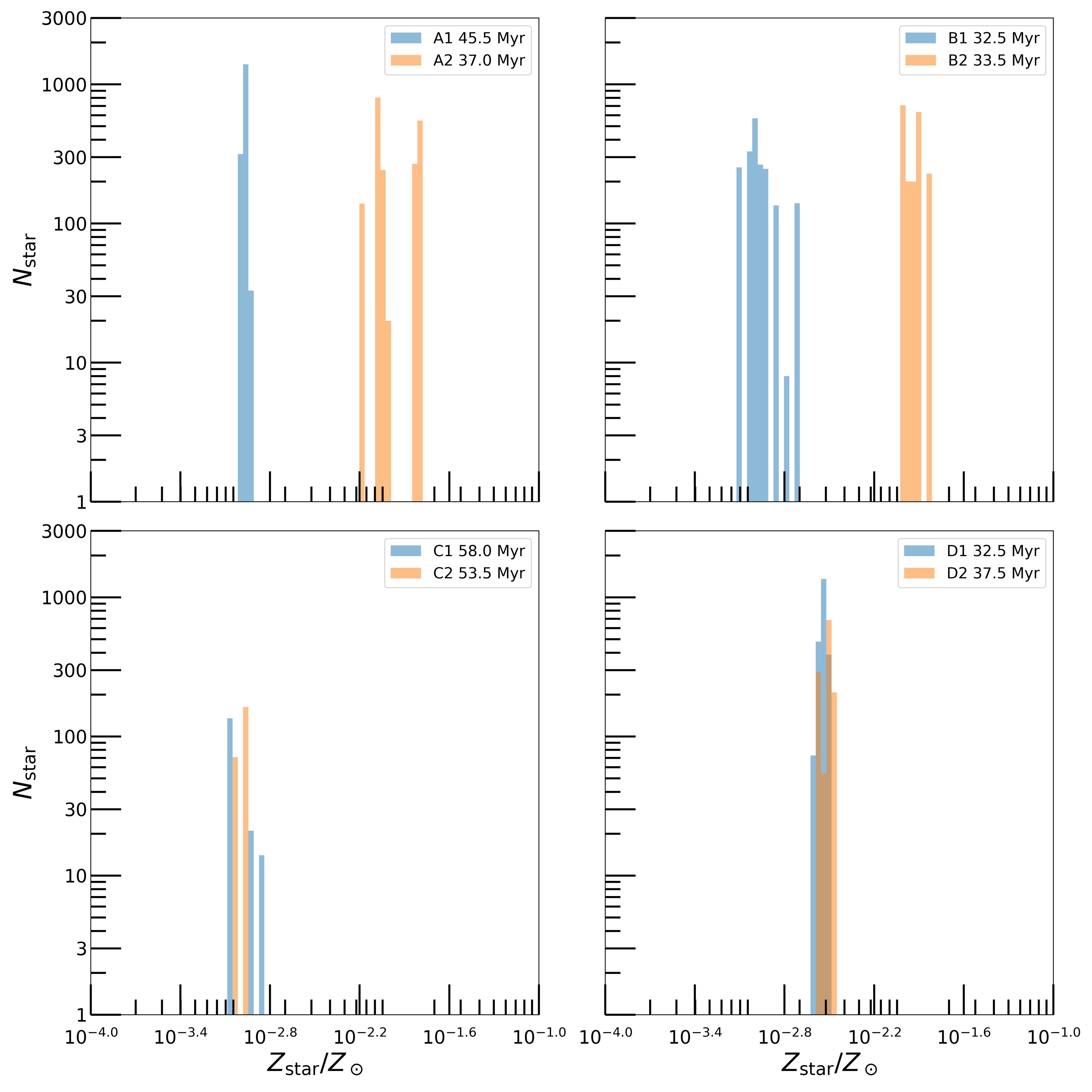}
	\caption{ Metallicity distribution of the Pop~II stars in first galaxies. Each model has a stellar population within a metallicity range of $ 10^{-3}- 2\times 10^{-2} \Zs$. Because only A and B models form the Pop~III stars, they create a more dispersed stellar metallicity distribution between A1 and A2, and B1 and B2 due to the metals from the Pop~III SNe. }
	\label{fig:dstellarmetal}
\end{figure}

We now examine the stellar metallicity of Pop~II stars by exhibiting their metallicity distributions in Figure~\ref{fig:dstellarmetal}.  Stars in A2 and B2 runs have the highest metallicities because new Pop~III stars die as PI~SNe and inject the most metals into the halo center. Overall, the B and D models have higher metallicities than the A and C models because of the initial PI~SNRs. The metallicity distribution of the C and D series is more confined than the A or B series because no Pop~III stars form to further enrich the gas. Small variations in metallicity within C and D series are due to the variations in the initial distributions of SNRs. 
The gas at the center of the halo in the D series is more turbulently mixed by collisions between PI~SNRs there (as seen in the (d) panel in Figure~\ref{fig:rgaszooma}). Therefore, there is even less variation in metallicity than in the D series. 

\subsection{Stellar Dynamics}

\begin{figure*}
\centering
\includegraphics[width=\textwidth]{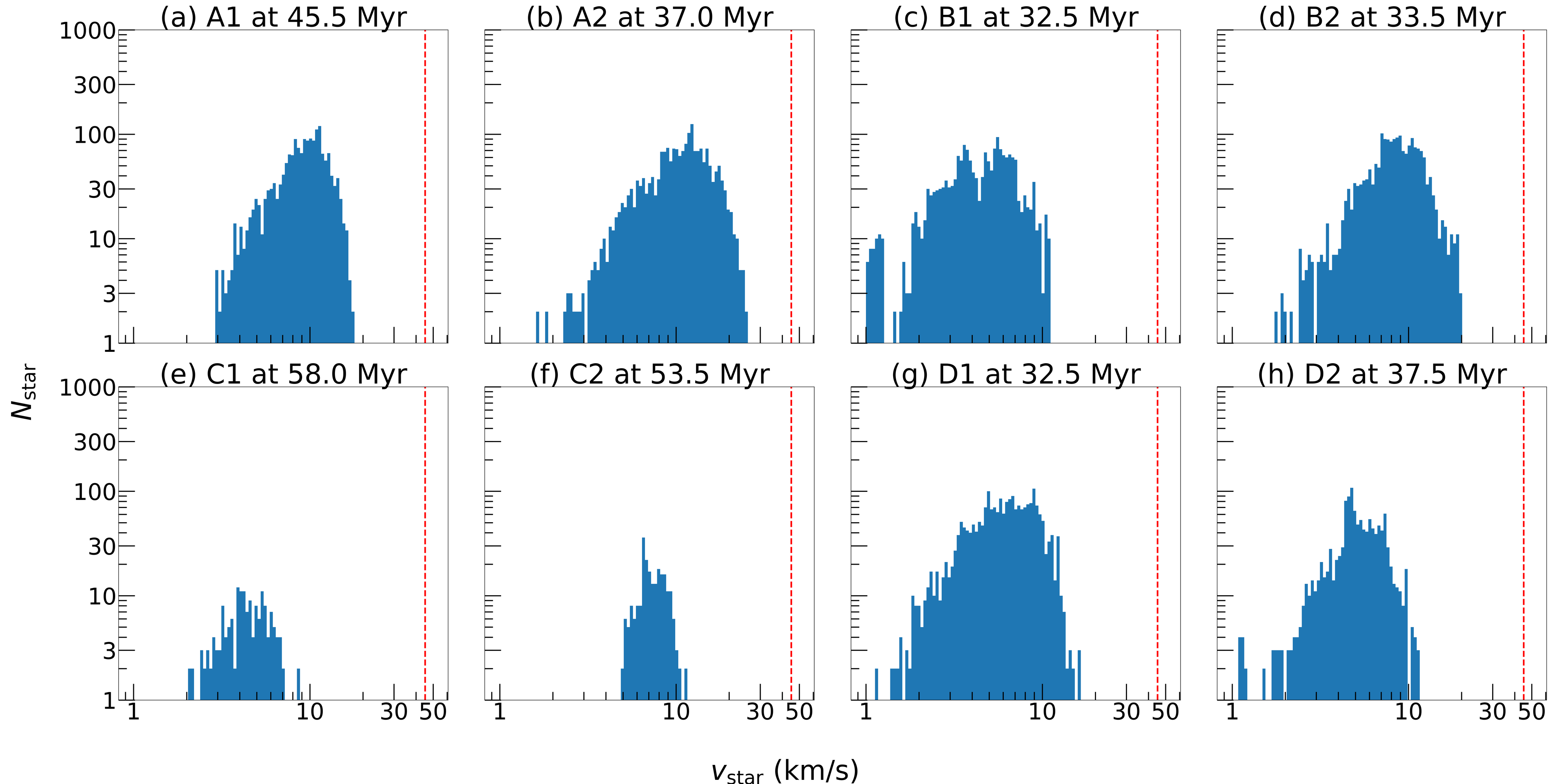}
\caption{Stellar velocity distributions at the end of each run. In all models, the drift velocities of Pop~II stars are below the escape velocity of halos of $\sim 45$ km s$^{-1}$ denoted with red lines.} 
\label{fig:rsv}
\end{figure*}

Newly born Pop~II stars can inherit a drift velocity from the motion of their star-forming cloud. If this velocity exceeds the escape velocity of a halo, stars will escape the halo without contributing to its luminosity. We examine this probability of runaway Pop~II stars by showing the final Pop~II star velocity distributions in Figure~\ref{fig:rsv}.  Pop~II stars have velocities of $< 30$ km s$^{-1}$, but the escape velocity of a $10^9 \Ms$ halo is $\sim  45$ km s$^{-1}$ in A - D. Therefore, all Pop~II stars will remain in their host halo. 
 
\subsection{Correlating the Pop~III IMF with the First Galaxies}

Most Pop~II stars in the first galaxies form with  $Z \geq 10^{-3} \Zs$, exceeding the metallicity of extremely metal-poor (EMP) stars, $Z \sim 10^{-5} - 10^{-4} \Zs$ \citep{Umeda&Nomoto2002,frebel10}.   \rev{Consequently, EMP stars are not representative of Pop~II stars in the earliest galaxies.} Their mass function follows a power law whose indices depend on SNR type and the Pop~III IMF.  Top--heavy Pop III IMFs \citep{hir15,hir17} create metal-rich PI SNRs that enhance metal cooling and eventually lead to steeper slopes for the early Pop~II IMF, less massive Pop~II stars, and dimmer galaxies at early times than Salpeter Pop III IMFs.  

\section{Conclusion}
\label{Conclusion}

We \rev{have investigated how Pop~III SNRs bridge the death of the first stars and the rise of the first galaxies with high-resolution simulations.  We find that that they rapidly enriched gas in massive halos with metals and drove turbulence as they gravitated toward the centers of the halos.  Subsequent populations of stars appeared at earlier times in these galaxies if the Pop~III IMF was top-heavy because the gas in halos became contaminated by metals more quickly, cooling and forming stars.  This trend is primarily due to PI~SNe producing nearly 100 times the metals of CC~SNe and more than 10 times the metals of HNe \citep{hw02,hw10}.}

PI~SNe in our protogalaxies consistently enrich star-forming gas to metallicities of 10$^{-3} - 10^{-2}$ \Zs, \rev{above those targeted by most surveys of EMP stars to date \citep[e.g.,][]{Cayrel2004,bc05,aoki14,plac16}.  This likely accounts for their failure to conclusively identify the 'odd-even' nucleosynthetic fingerprint of these explosions \citep{hw02,karl08}.  Indeed, in the first confirmed detection of the odd-even effect in a metal-poor star, LAMOST J1010+2358 \citep{xing23}, happened at a metallicity of [Fe/H] = -2.4, consistent with those of the galaxies in our models with PI SNRs.}

Our simulations also \rev{indicate that Pop~III stars were not a major component of most early galaxies} because gas in massive halos was usually polluted by metals from other Pop~III SNe during hierarchical assembly before it could collapse into pristine stars. In future work, we will evolve our galaxies to larger masses to better determine the evolution of the mass function of Pop~II stars over time.  Followups to the {\em JWST} CEERS and JADES surveys in coming years will achieve greater depths and probe the properties of the first galaxies and and second generation stars in the coming decade.


$$\rm Acknowledgements$$

The authors thank Li-Hsin Chen for her support with the \enzo\ simulations and analysis. We also thank Hiroyuki Hirashita and You-Hua Chu for useful comments. K. C. was supported by the National Science and Technology Council, Taiwan under grant no. MOST 110-2112-M-001-068-MY3 and the Academia Sinica, Taiwan under a career development award under grant no. AS-CDA-111-M04. Numerical simulations were performed at the National Energy Research Scientific Computing Center (NERSC), a U.S. Department of Energy Office of Science User Facility operated under Contract No. DE-AC02-05CH11231, at the Center for Computational Astrophysics (CfCA) at the National Astronomical Observatory of Japan (NAOJ), and at the TIARA Cluster at the Academia Sinica Institute of Astronomy and Astrophysics (ASIAA). 




\end{document}